\documentclass[%
 reprint,
 amsmath,amssymb,
 aps
]{revtex4-2}

\usepackage{graphicx}
\usepackage{dcolumn}
\usepackage{bm}
\usepackage{xcolor}
 \usepackage{ulem}
\newcommand{\sign}{{\rm sign}}
\newcommand{\ii}{{\rm i}}
\newcommand{\yo}[1]{{\color{black} #1}}

\definecolor{dark-green}{RGB}{0, 128, 0}

\begin{document}

\preprint{APS/123-QED}

\title{Bulk-Edge Correspondence Recovered in Incompressible Geophysical Flows}

\author{Yohei Onuki}
\email{onuki@riam.kyushu-u.ac.jp}
\affiliation{Research Institute for Applied Mechanics, Kyushu University, 6-1 Kasuga-koen, Kasuga, Fukuoka, Japan}%

\author{Antoine Venaille}
\author{Pierre Delplace}
\affiliation{ENS de Lyon, CNRS, Laboratoire de Physique UMR5672, F-69342 Lyon, France}%

\date{\today}

\begin{abstract}
Bulk-edge correspondence is a cornerstone in topological physics, establishing a connection between the number of unidirectional edge modes \yo{in physical space and a Chern number, an integer that counts phase singularities of the eigenmodes in parameter space.} In continuous media, violation of this correspondence has been reported \yo{when some of the frequency wavebands are unbounded}, resulting in weak topological protection of chiral edge states. Here, we propose a strategy to reestablish strong bulk-edge correspondence in incompressible \yo{rotating stratified flows, taking advantage of a natural cutoff frequency provided by density stratification. The key idea involves the introduction of an auxiliary field  to handle the divergence-free constraint.} This approach highlights the resilience of \yo{internal} coastal Kelvin waves near vertical walls under varying boundary conditions.
\end{abstract}

\maketitle

\section{Introduction}
Inspired by pioneering work in the context of the quantum Hall effect \cite{thouless1982quantized,hatsugai1993chern}, recent years have witnessed tremendous utility of topology to elucidate the emergence of unidirectional trapped modes in continuous media. This approach led to significant advances in understanding geophysical flows \cite{delplace2017topological,perrot2019topological,venaille2021wave}, stellar dynamics \cite{perez2022unidirectional,leclerc2022topological}, active matter \cite{shankar2017topological,shankar2022topological}, plasma physics \cite{parker2020topological,parker2020nontrivial,parker2021topological,fu2021topological,fu2022dispersion,qin2023topological}, and its application range is further expanding \cite{finnigan2022equatorial,li2023equatorial,zhu2023topology,green2020topological}. The basic concept is to compute a topological invariant named the Chern number for each wave band of a simple bulk problem admitting plane wave solutions. This number counts \yo{phase singularities of the eigenmodes in wavenumber space} and can then be used to predict the wave spectrum in more complicated situations involving, for instance, boundaries or spatially varying parameters. Because waves of topological origin are robust to continuous changes in the system's properties, they are thought to play exceptional roles in energetics and transport phenomena. In the case of continuous media with spatially varying parameters, the existence of unidirectional modes is guaranteed by index theorems \cite{faure2023manifestation,delplace2022berry,venaille2023ray}. By contrast, topological protection of the edge states along boundaries of a continuous system remains an open question, owing to the lack of compactness in reciprocal space unlike lattice systems holding Brillouin zones \cite{silveirinha2015chern,silveirinha2016bulk}. Moreover, even when one somehow regularizes eigenmodes to compactify wave number space, the resulting Chern numbers do not correctly predict the number of edge modes \cite{tauber2019bulk,tauber2020anomalous,tauber2023topology}. Up to now, this violation of bulk-edge correspondence has been a major limitation of the topological approach to continuous media.

In these prior studies, the main obstacle to establishing a bulk-edge correspondence arose from the presence of wavebands with unbounded frequencies, akin to sound waves. It is then natural to expect this issue to be solved in a model equipped with a high-frequency cutoff. Indeed, the present study demonstrates that a bulk-edge correspondence is recovered in the low-frequency part of the geophysical flow system, i.e., rotating stratified fluid with the incompressible condition imposed to effectively filter out acoustic waves. This achievement is reached \yo{with the help} of a noticeable theoretical tool: we introduce an additional degree of freedom referred to as an \textit{auxiliary field} that replaces a divergence-free constraint with the existence of a stationary divergent flow. Thanks to this spurious mode, the original wave problem transforms into a Schrödinger-like formulation featuring a Hermitian wave operator, with each waveband exhibiting an eigenmode bundle in compact wave number space. In contrast to earlier work, a series of procedures now eliminate the need to introduce regularization terms \cite{tauber2020anomalous} or replace boundaries with an interface problem when defining a Chern number for each waveband \cite{venaille2021wave}. We establish the correspondence between those bulk Chern numbers and the existence of unidirectional edge modes trapped along the system's lateral boundaries. We identify these edge modes as internal coastal Kelvin waves and, based on Levinson's theorem \cite{graf2013bulk,tauber2023topology}, verify their robustness against changes in boundary conditions.

\section{Model equations}
We consider a linear inviscid model of rotating and stratified fluid under the Boussinesq and traditional approximations~\cite{vallis2017atmospheric},
\begin{subequations} \label{eq:basic}
\begin{eqnarray}
\partial_t \bm{u} & = & - f \hat{\bm{z}} \times \bm{u} + N \theta \hat{\bm{z}} - \nabla p \\
\partial_t \theta & = & - N w .
\end{eqnarray}    
\end{subequations}
The variables, $\bm{u} = (u,v,w)$ and $\theta$, represent the 3d velocity vector and the scaled buoyancy perturbation, respectively, which are functions of space $\bm{x} = (x, y, z)$ and time $t$. We let $\hat{\bm{z}}$ denote a unit vector pointing upwards, fix the buoyancy frequency $N>0$, and allow the Coriolis parameter $f$ to be a function of~$\bm{x}$. Pressure $p$ is a Lagrange multiplier that associates the divergence-free constraint, $\nabla \cdot \bm{u} = 0$, thus satisfying $\nabla^2 p = - \nabla \cdot (f \hat{\bm{z}} \times \bm{u}) + N \partial_z \theta$. \yo{The derivation of \eqref{eq:basic} from a more fundamental model is presented in Appendix~\ref{app:model_equation}.} The present system conserves energy, $(1/2) \int (\vert \bm{u} \vert^2 + \theta^2) d\bm{x}$, under a suitable boundary condition. Accordingly, any solution should be decomposed into a set of stationary and oscillatory modes, whose frequency property is the scope of our interest.

\begin{figure}[t]
\includegraphics[bb=0 0 567 280, width=0.8\columnwidth]{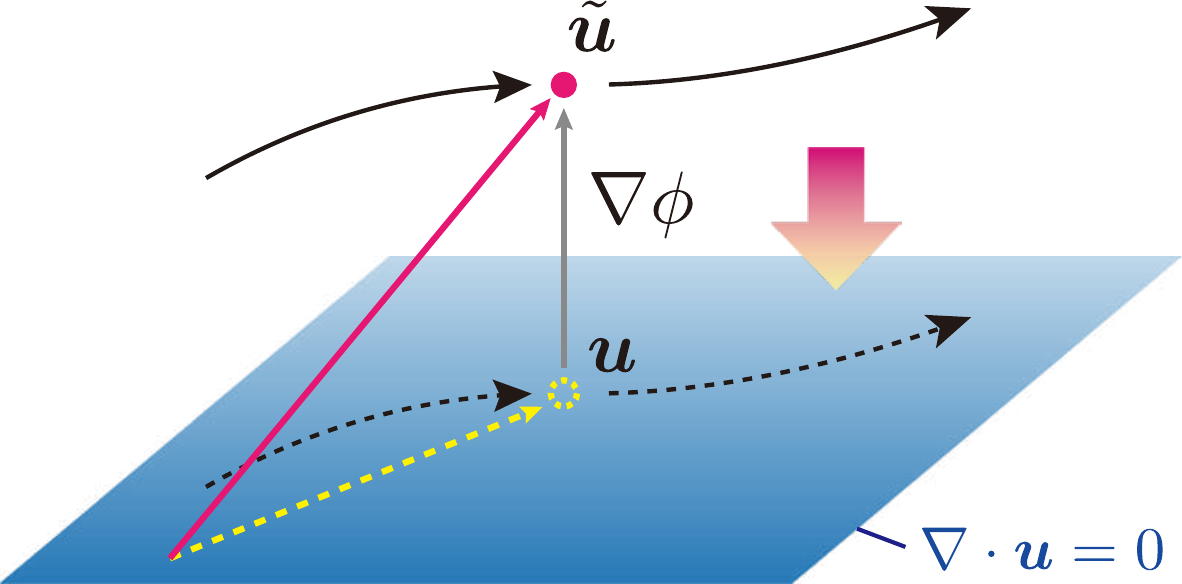}
\caption{\label{fig:projection_operator} Schematic illustration of the auxiliary field. We combine an original velocity vector $\bm{u}$ with the gradient of an additional variable $\phi$ to compose an extended velocity vector~$\tilde{\bm{u}}$ \yo{(combination of the yellow dotted and gray solid arrows produces the red solid arrow)}. Conversely, we can recover $\bm{u}$ from $\tilde{\bm{u}}$ at any instance by projection onto the subspace specified by the incompressible constraint $\nabla \cdot \bm{u} = 0$ \yo{(the thick arrow visualizes this projection)}. Trajectories of $\tilde{\bm{u}}$ and $\bm{u}$ in phase space\yo{, denoted by the solid and dotted curves, respectively,} are parallel as the condition $\partial_t \tilde{\bm{u}} = \partial_t \bm{u}$ identically holds.}
\end{figure}

Even though \eqref{eq:basic} holds four variables, $u, v, w$ and $\theta$, the static constraint for the velocity field, $\nabla \cdot \bm{u} = 0$, reduces the genuine degrees of freedom to three, which forbids the direct use of well-established topological machinery. To circumvent this difficulty, we shall introduce a new variable $\phi(\bm{x})$, an arbitrary function that does not depend on time, and define an extended velocity vector as $\tilde{\bm{u}} = (\tilde{u}, \tilde{v}, \tilde{w}) = \bm{u} + \nabla \phi$. Importantly, identifying $\tilde{\bm{u}}$ is equivalent to identifying the set of variables $(\bm{u}, \phi)$; for a given $\tilde{\bm{u}}$, its compressible part is determined by solving $\nabla^2 \phi = \nabla \cdot \tilde{\bm{u}}$ with a suitable boundary condition, and the incompressible part is the residual $\bm{u} = \tilde{\bm{u}} - \nabla \phi$. This relationship is shown schematically in Fig.~\ref{fig:projection_operator}.

The extended state vector, $\bm{\psi} = (\tilde{\bm{u}}, \theta)$, now recovers the four degrees of freedom thanks to the auxiliary field. Because $\partial_t \bm{u} = \partial_t \tilde{\bm{u}}$ always holds, the left-hand sides of \eqref{eq:basic} are identified as $\partial_t \bm{\psi}$, and the right-hand sides are at any instance computable from $\bm{\psi}$. \yo{Accordingly, the governing equations transform into a set of prognostic and diagnostic equations,
\begin{align*}
\partial_t \tilde{\bm{u}} & = - f \hat{\bm{z}} \times (\tilde{\bm{u}} - \nabla \phi) + N \theta \hat{\bm{z}} - \nabla p \\
\partial_t \theta & = - N (\tilde{w} - \partial_z \phi) \\
\nabla^2 p & = - \nabla \cdot \left( f \hat{\bm{z}} \times (\tilde{\bm{u}} - \nabla \phi) \right) + N \partial_z \theta \\
\nabla^2 \phi & = \nabla \cdot \tilde{\bm{u}} .
\end{align*}
We can write these equations symbolically as $\partial_t \bm{\psi} = \mathcal{L} \bm{\psi}$, where $\mathcal{L}$ is a linear integral operator. The precise form of $\mathcal{L}$ is specified by the boundary conditions of $p$ and $\phi$ and generally not easy to write down. In this study, we assume that this boundary condition does not violate energy conservation, so that $d ( \bm{\psi}, \bm{\psi} ) / dt = 0$ holds for a certain inner product in real vector space. If the system is unbounded, we readily verify that $( \bm{\psi}, \bm{\psi} ) = (1/2) \int (\vert \tilde{\bm{u}} \vert^2 + \theta^2) d \bm{x} = (1/2) \int (\vert \bm{u} \vert^2 + \vert \nabla \phi \vert^2 + \theta^2) d \bm{x}$ fulfills this requirement. When there exists a boundary, we may extend the definition of inner product to include energy stored in the surface (Section~\ref{sec:wall}). In any case, energy conservation ensures the skew symmetry of $\mathcal{L}$, which, in turn, defines a Hermitian operator in complex space as $\mathcal{H} \equiv \ii \mathcal{L}$. Consequently, we obtain a Schr\"odinger-like equation
\begin{align} \label{eq:schrodinger}
\ii \partial_t \bm{\psi} = \mathcal{H} \bm{\psi} .
\end{align}}
The extended model for $\bm{\psi}$ differs from the original system \eqref{eq:basic} in that the new solution includes a spurious mode, $\bm{\psi} = (\nabla \phi, 0)$. However, the two models are identical in their dynamic parts, since the auxiliary field is stationary in time. In other words, the finite part of the spectrum of $\mathcal{H}$ coincides with the set of natural frequencies in the oscillatory solutions of the original equation \eqref{eq:basic}. Therefore, the investigation of $\mathcal{H}$ is enough in the present discussion.

\section{Bulk problem} 
\begin{figure}[b]
\includegraphics[bb=0 0 1035 827, width=\columnwidth]{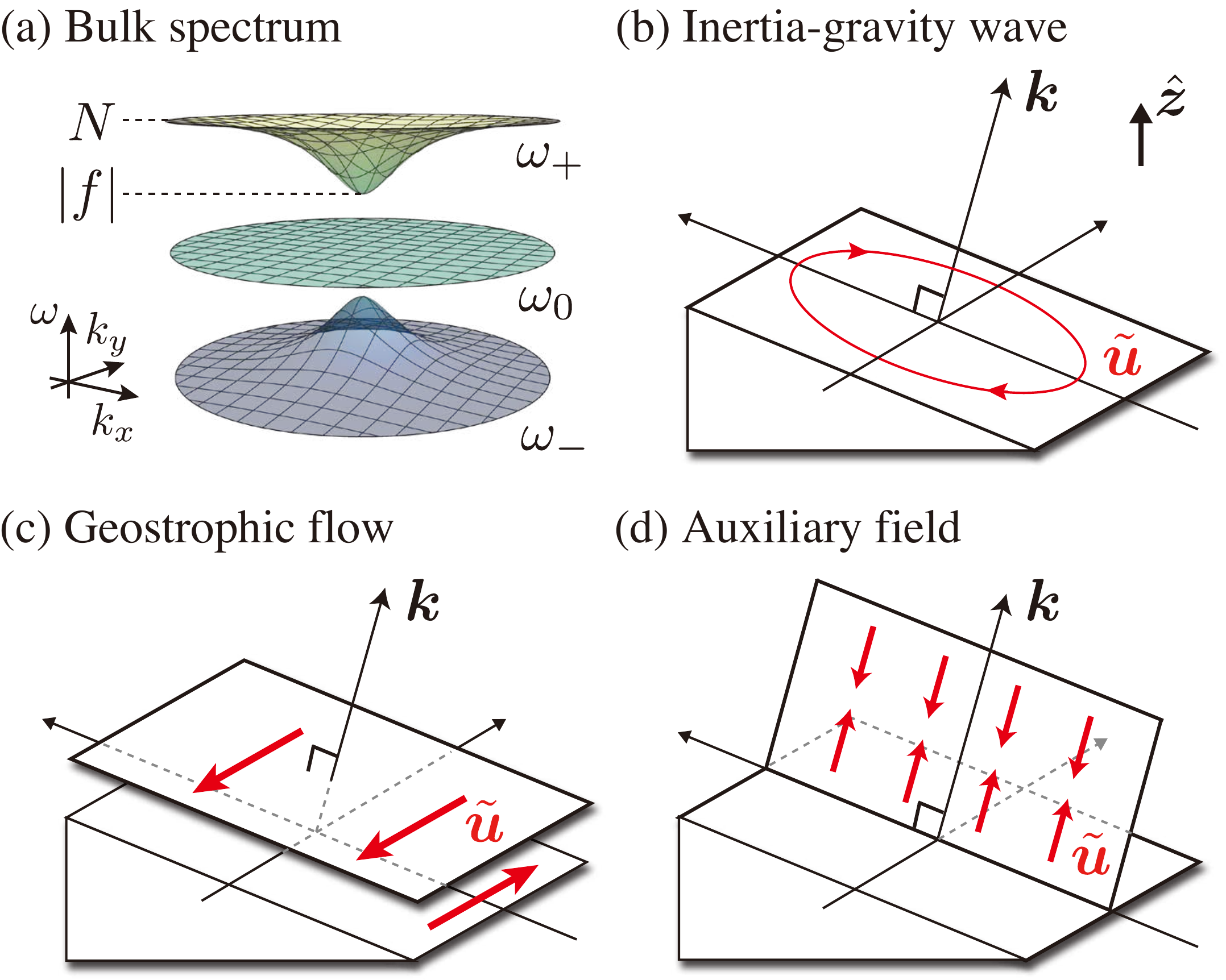}
\caption{\label{fig:polarization} (a) Frequency spectrum of the bulk states against the horizontal wave numbers. The middle band, $\omega = \omega_0$, is two-fold degenerate. (b-d) Polarization relations of the bulk eigenstates are illustrated by drawing flow trajectories associated with the extended velocity vector $\tilde{\bm{u}}$.}
\end{figure}

Fixing $f$ and ignoring the boundary condition allow a solution in the form of $\bm{\psi} = \hat{\bm{\psi}} e^{\ii (\bm{k} \cdot \bm{x} - \omega t)}$. Here, $\hat{\bm{\psi}}$ is a normalized eigenvector satisfying $\omega \hat{\bm{\psi}} = \mathcal{H}_k \hat{\bm{\psi}}$, with the Hermitian matrix $\mathcal{H}_k$ represented by
\begin{align*}
\frac{\ii}{\vert \bm{k} \vert^2} \begin{pmatrix}
0 & k_z^2 f & - k_y k_z f & - k_x k_z N \\
- k_z^2 f & 0 & k_x k_z f & - k_y k_z N \\ k_y k_z f & - k_x k_z f & 0 & (k_x^2 + k_y^2) N \\ k_x k_z N & k_y k_z N & - (k_x^2 + k_y^2) N & 0
\end{pmatrix} ,
\end{align*}
where $\bm{k}=(k_x, k_y, k_z)$. The four eigenvalues read $\{ \omega_-, \omega_0, \omega_0, \omega_+ \}$ with $\omega_\pm = \pm \sqrt{(k_x^2 + k_y^2)N^2 + k_z^2 f^2} / \vert \bm{k} \vert$ and $\omega_0 = 0$. The finite parts, $\omega_\pm$, represent the dispersion relations of inertia-gravity waves. The zero-frequency parts involve the geostrophic flow, which is a genuine solution of the original model, and the auxiliary field, namely, a spurious mode. In this problem, an eigenvector represents the polarization relations for each mode, as illustrated in Fig.~\ref{fig:polarization}.

Since the eigenvectors are now parameterized by $\bm{k}$ and $f$, they compose fiber bundles over a certain manifold. As reported in \cite{delplace2022berry,faure2023manifestation,venaille2023ray}, a mathematically accomplished way to define such a bundle exhibiting non-trivial topology is considering a closed surface in 3d space spanned by two components in the wave vector and one spatial parameter. Chern numbers on this surface predict the number of interface-trapped modes whose dispersion curves connect separate frequency bands. In the present case, on the other hand, we obtain a more powerful result by investigating fiber bundles in pure wave number space. We shall fix $k_z$ and $f$ and focus on a plane spanned by the horizontal wave numbers, $k_x$ and $k_y$. Eigenvectors are originally defined in an open domain, $(k_x, k_y) \in \mathbb{R}^2$, but now we regard it as a complex plane composed of $k \equiv k_x + \ii k_y$ and identify the infinity, $k = \infty$, to define a Riemann sphere, $S^2$. Thus, we derive a set of fiber bundles over a compactified surface. Note that this procedure makes sense only when the linear space spanned by eigenvectors in each frequency band is identical at $k = \infty$ \cite{tauber2019bulk,souslov2019topological}. Fortunately, the present system satisfies this requirement as the inertia-gravity wave converges to a vertical buoyancy oscillation in the large horizontal wave number limit. Although the eigenvector of the geostrophic mode alone does not converge in this limit, since it degenerates with the auxiliary field, the combined eigenspace is consistently defined throughout~$S^2$. \yo{See Appendix~\ref{app:eigenbundle} for the detailed structure of these bundles.}

Now, we have three fiber bundles over $S^2$, with structure groups $U(1)$, $U(2)$ and $U(1)$, corresponding to the frequency bands, $\omega_-$, $\omega_0$ and $\omega_+$, respectively. The Chern number for each band is computed via the formula
\begin{align} \label{eq:chern}
\mathcal{C}_n = \frac{1}{2 \pi} \int_{S^2} \textrm{tr} B_n dk_x dk_y  \quad \text{with} \quad n \in \{ -, 0, + \},
\end{align}
where $B_n$ is the Berry curvature, whose precise definition is found, e.g., in \cite{vanderbilt2018berry}. Generally, $B_n$ is a matrix, but for the simplest cases of $U(1)$-bundles, it becomes a scalar and is conveniently computed by $B_n = \ii ( \langle \partial_{k_x} \hat{\bm{\psi}}, \partial_{k_y} \hat{\bm{\psi}} \rangle - \langle \partial_{k_y} \hat{\bm{\psi}}, \partial_{k_x} \hat{\bm{\psi}} \rangle )$, where brackets denote the inner product of complex vectors. \yo{The direct evaluation of \eqref{eq:chern} then} yields $\mathcal{C}_- = - f / \vert f \vert$ and $\mathcal{C}_+ = f / \vert f \vert$ for inertia-gravity wave bands \yo{(Appendix~\ref{app:eigenbundle})}. For the zero-frequency band, the computation of $B_n$ is involved. We here employ a shortcut; since the Chern numbers summed up over all the frequency bands always become 0, $\mathcal{C}_0 = 0$ derives without analytical computation. The summarized results are\yo{
\begin{align} \label{eq:chern_number}
\{ \mathcal{C}_-, \mathcal{C}_0, \mathcal{C}_+ \} = \begin{cases}
 \{ - 1, 0, 1 \} & \text{for} \quad f > 0 \\
\{ 1, 0, - 1 \} & \text{for} \quad f < 0 .
\end{cases}    
\end{align}
The changes in the Chern numbers dependent on the sign of $f$ imply the existence of topologically protected states along the equator. This point was discussed in \cite{tauber2019bulk} and the consistency of the present result with the earlier work is checked in Appendix~\ref{app:equator}.
}


\section{Wall states} \label{sec:wall}
\subsection*{General criteria of Hermiticity}
An advantage in the Chern numbers defined purely in wave number space is that it predicts the number of edge states not only around an interface within the system but also along an external boundary. We shall inspect this nature in a particular situation: $f$ is again fixed, but a lateral wall restricts the fluid motion in the meridional direction. \yo{Let $V = \{ \bm{x} = (x, y, z) \in \mathbb{R}^3 \vert y \geq 0 \}$ be the domain of the fluid. On the domain wall $\partial V$, we impose $\phi = 0$, which acts as the boundary condition of the Poisson equation determining the auxiliary field, $\nabla^2 \phi = \nabla \cdot \tilde{\bm{u}}$. The energy equation of the fluid motion is derived as
\begin{align*}
\frac{d}{dt} \int_V \frac{\vert \tilde{\bm{u}} \vert^2 + \vert \theta \vert^2}{2} d \bm{x} + \int_{\partial V} \left( - \frac{p v^\dag + c.c.}{2} \right) dx dz = 0 .
\end{align*}
The system is Hermitian regarding the energy norm if either (i) the surface integration term identically vanishes or (ii) the surface integration term turns into the temporal integration of a positive definite functional that corresponds to the energy stored in the boundary.

\subsection*{Elastic boundary condition}
To construct a physically relevant model, we shall assume that elements of the boundary wall are made from elastic material (continuously aligned infinitesimal springs as illustrated in Fig.~\ref{fig:edge_modes}a). We write the displacement of the surface element as $\xi$, and write the equation of motion as
\begin{align} \label{eq:bc_general}
\sigma \ddot{\xi} + \lambda \xi = - p .
\end{align}
Here, $\sigma \geq 0$ is the spring mass density per unit area, and $\lambda > 0$ is the spring constant. Surface displacement motion is related to the fluid velocity by $\dot{\xi} = v$. The Hermitian condition (ii) raised above is then fulfilled with the modified energy norm,
\begin{align} \label{eq:norm_elastic}
( \bm{\psi}, \bm{\psi} ) = \int_V \frac{\vert \tilde{\bm{u}} \vert^2 + \vert \theta \vert^2}{2} d \bm{x} + \int_{\partial V} \frac{\sigma \vert v \vert^2 + \lambda \vert \xi \vert^2}{2} dx dz ,
\end{align}
where we regard $\xi$ as part of the state vector, $\bm{\psi}$.

In the following, for simplicity, we mainly consider massless springs, $\sigma = 0$, in which the elastic force is always equilibrated with the pressure force. As a result, by setting $a = 1 / \lambda$ and taking the temporal derivative of \eqref{eq:bc_general}, we derive a reduced form of the boundary condition as
\begin{align} \label{eq:boundary_y0}
v = - a \partial_t p \quad \text{and} \quad \phi = 0 \quad \text{at} \quad y = 0 .
\end{align}
In this particular limit, the energy norm \eqref{eq:norm_elastic} is replaced by
\begin{align}
( \bm{\psi}, \bm{\psi} ) = \int_V \frac{\vert \tilde{\bm{u}} \vert^2 + \vert \theta \vert^2}{2} d\bm{x} + \int_{\partial V} \frac{a \vert p \vert^2}{2} dx dz .
\end{align}
This approximation is reasonable when considering an oscillatory solution whose frequency is much lower than the natural frequency of the spring, $\omega_s \equiv \sqrt{\lambda / \sigma}$. However, effect from the inertia term in \eqref{eq:bc_general} neglected in the present model should be taken into account for the special case of the very high-frequency modes. We elucidate this limit in Appendix~\ref{app:high_frequency}. Now, the boundary condition is specified by a single parameter $a \geq 0$.} From a geophysical viewpoint, the rigid wall condition, $a = 0$, is typical, but here we vary $a$ to assess the resilience of the edge states, as expected from topological protection.

\subsection*{Bulk-edge correspondence}
We shall seek an eigenstate in the form of $\bm{\psi} = \bm{\Psi}(y) e^{\ii (k_x x + k_z z - \omega t)}$ with $\omega \neq 0$. For this oscillatory solution, the auxiliary field strictly vanishes. We then derive a single equation for pressure,
\yo{
\begin{align}
(N^2 - \omega^2) p_{yy} - [(N^2 - \omega^2) k_x^2 - (\omega^2 - f^2) k_z^2] p = 0 ,
\end{align}
whose solution is generally represented by $p = A_- e^{- \ii \kappa y} + A_+ e^{\ii \kappa y}$ with
\begin{align} \label{eq:kappa}
\kappa = \sqrt{\frac{(\omega^2 - f^2) k_z^2 - (N^2 - \omega^2) k_x^2}{N^2 - \omega^2}} .
\end{align}
}
The meridional wave number $\kappa$ is real if and only if $\omega$ belongs to the bulk spectrum. Otherwise, $\kappa$ is a pure imaginary number satisfying $\Im \kappa \geq 0$. Then, for the solution to be finite for $y \to \infty$, $A_- = 0$ must be satisfied, which, combined with the boundary condition, determines the dispersion relations of wall-trapped modes. \yo{The explicit form of the dispersion relations are derived in Appendix~\ref{app:edge_mode}.}

\begin{figure*}
\includegraphics[bb=0 0 1097 389, width=\textwidth]{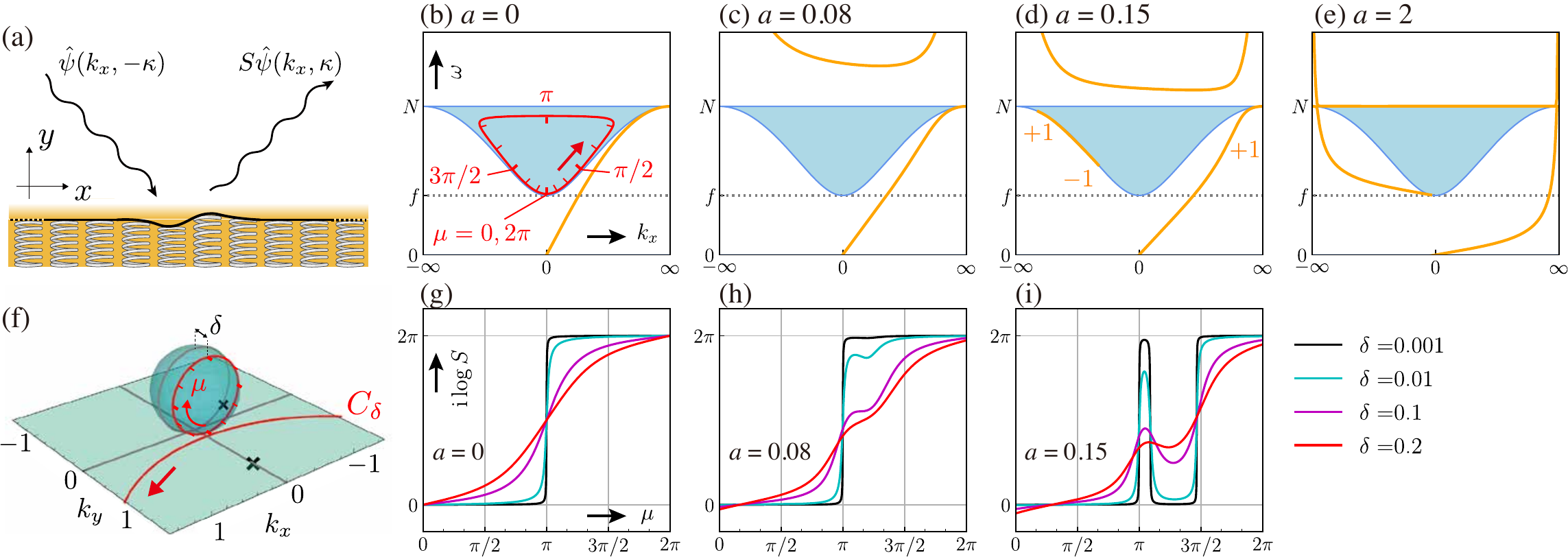}
\caption{\label{fig:edge_modes} Edge states on a lateral boundary; as illustrated in (a), a plane elastic wall is placed at $y = 0$. (b-e)~Plots of edge-mode dispersion curves (orange) for several $a$ with the positive inertia-gravity wave band (cyan). The scale of the horizontal axes is $\arctan(k_x / k_z)$, which contracts $-\infty \leq k_x \leq \infty$ to a finite range. (f)~Closed paths $C_\delta$ in the wave number plane are defined as $(k_x, k_y) = (\sqrt{1/4 - \delta^2} \sin \mu, \delta) / (1/2 + \sqrt{1/4 - \delta^2} \cos \mu)$ with $0 \leq \mu \leq 2 \pi$, and a particular case of $\delta = 0.2$ and its projection onto the Riemann sphere are depicted. The inertia-gravity wave frequency on this path is represented by a red curve in (b). (g-i)~Plots of phases in the scattering coefficient $S$ for positive inertia-gravity waves, computed along $C_\delta$ for different $a$ as corresponding to (b-d). Throughout the analysis, $(k_z, f, N) = (2, 2, 5)$ is chosen, and the gauge of the bulk eigenvector is fixed such that a singular point is at $k = \ii$, as denoted by crosses in (f). See \yo{Appendices~\ref{app:edge_mode} and \ref{app:scatter}} for more details.} 
\end{figure*}

Several results from algebraic computations are demonstrated in Fig.~\ref{fig:edge_modes}. There exists a single trapped mode in the rigid wall case, $a = 0$. Its dispersion curve coincides with that of non-rotating internal gravity waves, the salient feature of the coastal Kelvin wave. As $a$ increases, this Kelvin wave structure is modified, and new trapped modes appear. We first find a high-frequency mode whose dispersion curve goes downwards approaching $N$. This mode originates from the elastic force from the wall. When $a$ is small, its dispersion relation coincides with the natural frequency of massless springs dressed in fluid with finite mass. \yo{This property is elucidated in Appendix~\ref{app:high_frequency}}. In the large $a$ limit, it converges to the buoyancy oscillation while maintaining asymptotically high frequencies around $k_x = \pm \infty$. When $a$ surpasses a certain threshold, another mode emerges from the low-frequency side of the inertia-gravity wave band, and it approaches $f$, still possessing two connection points to the bulk spectrum. This mode partly resembles an edge wave identified in a shallow-water system with an abrupt change in water depth \cite{iga1995transition}. In the present model, however, this edge mode does not play any role in filling bulk gaps, in contrast to the shallow-water case in which its dispersion curve connects the topographic Rossby and inertia-gravity wave bands \cite{venaille2021wave}. The Kelvin wave dispersion curve also goes downward as $a$ increases, but it keeps connecting the separate bulk bands at $(k_x, \omega) = (0, 0)$ and $(k_x, \omega) = (\infty, N)$. Therefore, there always exists a unique wall state in the bulk gap between $\omega = 0$ and $\omega = f$. Moreover, it is a chiral mode because it always propagates in one direction with the wall on its right-hand side when $f > 0$.

Following \cite{tauber2020anomalous}, we assign indices to the wall-trapped modes by counting their connection points to the bulk bands in the spectrum. Specifically, let us count $-1$ when the edge-mode dispersion curve goes rightwards from a connection point and $+1$ when it goes leftwards. We then write $\mathcal{N}_n$ with $n \in \{ -, 0, + \}$ as the number of points counted over each frequency band. Because the spectrum is now point-symmetric with respect to the origin, we always hold \yo{$\mathcal{N}_0 = 0$} and $\mathcal{N}_+ = - \mathcal{N}_-$. Then, direct inspection of Fig.~\ref{fig:edge_modes} tells us $\{ \mathcal{N}_-, \mathcal{N}_0, \mathcal{N}_+ \} = \{ - 1, 0, 1 \}$ for any $a$. If we flip the sign of $f$, the overall spectrum is reflected around the $\omega$-axis, and $\mathcal{N}_n$ change their signs. We have thus confirmed the coincidence of the two sets of indices, $\mathcal{C}_n = \mathcal{N}_n$.

A concrete theoretical foundation underlying the present bulk-boundary correspondence is provided by Levinson's theorem \cite{tauber2020anomalous}. Its first step is to write down the eigenstates in the form of $\bm{\Psi} = \hat{\bm{\psi}} (k_x, -\kappa) e^{- \ii \kappa y} + S(k_x, \kappa) \hat{\bm{\psi}} (k_x, \kappa) e^{\ii \kappa y}$, where $\hat{\bm{\psi}}$ shares the functional dependence on $k_x$ and $\kappa$ with the eigenvector of the positive-frequency inertia-gravity wave solution, while the range of $\kappa$ is now extended to the complex plane. This expression represents wave reflection on the wall, in which $S$ acts as the scattering coefficient. Since we are dealing with a dissipationless process, if $\kappa$ is real, $S$ must belong to $U(1)$, representing the wave's phase shift upon reflection. More generally, $S$ is a complex function, and its pole corresponds to the edge-mode solution. Due to this singularity in $S$ on the edge-mode dispersion curves, in the bulk band region where $S \in U(1)$, it exhibits phase jumps at the connection points to the edge modes. We confirm this character by computing $S$ along a series of closed paths, $C_\delta$, on the Riemann sphere (\yo{see Appendix~\ref{app:scatter} for the detailed procedure to compute $S$}). As the path approaches the great circle representing the $k_x$-axis, it gets close to the margin of the inertia-gravity wave band (Fig.~\ref{fig:edge_modes}b, f). In this limit, the phase in $S$ exhibits a sharp staircase structure with jumps by $2 \pi$ or $- 2 \pi$ (Fig.~\ref{fig:edge_modes}g-i). Indeed, the locations of these jumps always coincide with the connection points from the bulk band to the edge-mode dispersion curves. Finally, the net increment in the phase over a path coincides with the winding number of a section of the inertia-gravity wave bundle, which is equivalent to the Chern number \cite{graf2021topology}. Consequently, we relate the two indices via the formula,
\begin{align} \label{eq:Levinson}
\mathcal{N}_+ = \frac{\ii}{2 \pi} \oint_{C_\delta} d \log S = \mathcal{C}_+ ,
\end{align}
a form of Levinson's theorem \cite{graf2013bulk}. \yo{The validity of this relationship is also checked with a series of boundary conditions other than the present elastic wall (Appendix~\ref{app:different_bc})}.

In fact, the perfect coincidence of $\mathcal{C}_n$ and $\mathcal{N}_n$ is in contrast to recent reports on the violation of \eqref{eq:Levinson} occurring in a wide range of continuous media. For a shallow water model equipped with odd viscosity, Tauber et al.~\cite{tauber2020anomalous} verified the second equality in \eqref{eq:Levinson} but identified anomaly in the first. Specifically, they pointed out that phase jumps in $S$ at asymptotically high frequencies cannot connect to edge modes in the gap. The present stratified model escapes this trouble because its bulk spectrum is bounded. Later, Graf et al.~\cite{graf2021topology} clarified that an evanescent mode entering a refection process impedes the correspondence between phase jumps in $S$ and the connection points. In general, an evanescent wave arises during a free-wave reflection if an equation involves differentiation higher than the second order. As seen here, an inviscid flow model does not involve an evanescent wave. Employing a physically realizable model that does not demand artificial regularization enables recovering the normal bulk-edge correspondence.

\section{Conclusions}


\yo{The anomalous bulk-edge correspondence commonly arising in continuous media challenges the increasing use of topology in, for instance, fluid and electromagnetic problems. The major cause of this anomaly is the unbounded parts of the bulk spectrum, which produce the ghost states that contribute to the topological indices but with no footprint on the edge-mode dispersion curves~\cite{tauber2020anomalous}. In contrast, a three-dimensional fluid system with stable stratification in a rotating frame, as investigated in this study, exhibits the perfect correspondence of the bulk and edge states. Through explicit computations of the bulk and edge solutions with various boundary conditions, we have concluded that this correspondence is guaranteed by Levinson’s theorem.}





\yo{A key feature of the flow model examined in this study is the absence of bulk waves above the frequency cutoff set by stratification. To ensure this property, we have considered an incompressible fluid, as otherwise sound waves could propagate at higher frequencies. Technically, this incompressibility constraint leads to a non-standard wave problem, complicating the computation of a Chern number. To overcome this challenge, we introduce a mathematical technique called the auxiliary field. This approach involves adding an unphysical variable to the original set of equations. The extended degree of freedom effectively removes the divergence-free condition from incompressible fluid flows, allowing the governing equations to be described in a Schrödinger-like form.  Once the auxiliary field is taken away from the output, the original wave problem is recovered. These procedures enable the construction of a complete set of fiber bundles in wave number space, while keeping the system essentially unchanged. Consequently, the Chern numbers can be computed, successfully predicting the existence of edge modes for any boundary condition.}

\yo{Importantly, this machinery is not limited to fluids, but holds value for any continuous medium with static constraints. For example, an electromagnetic field that involves the solenoidal magnetic vector falls into the same class. Therefore, photonics and plasma wave dynamics, both of which are active areas in topological physics, may benefit from this method.}

\yo{The topologically protected edge mode identified in the present study corresponds to} the coastal Kelvin wave and its relatives ubiquitous in the \yo{stably stratified} ocean. \yo{Our formulation establishes the theoretical foundation for the robust propagation of such internal Kelvin waves along a complex wall as experimentally confirmed by \cite{ren2021robust}.} Interestingly, \yo{a similar} wall state is found \yo{even under the unstable stratification} \cite{favier2020robust}, and its possible topological origin is discussed in a very recent study \cite{zhang2023non}. Bulk-edge correspondence in such a non-Hermitian continuous system remains to be explored.

\begin{acknowledgments}
We appreciate invaluable comments from two anonymous reviewers. This work was supported by JSPS Overseas Research Fellowship, as well as KAKENHI Grant JP20K14556. This work is also supported by the Collaborative Research Program of the Research Institute for Applied Mechanics, Kyushu University.
\end{acknowledgments}

\appendix
\yo{\section{Derivation of the model equations} \label{app:model_equation}
A basic equation system for rotating and stratified fluid flows under the Boussinesq approximation is found in standard textbooks of geophysical fluid dynamics. For example, Eq.~(2.108) of \cite{vallis2017atmospheric} provides
\begin{align*}
D_t \bm{u} + 2 \bm{\Omega} \times \bm{u} & = b \hat{\bm{z}} - \nabla p \\
\nabla \cdot \bm{u} & = 0 \\
D_t b & = 0 ,
\end{align*}
where symbols of velocity, pressure, and the unit vector pointing upwards are changed from those in the original text to $\bm{u} = (u, v, w)$, $p$, and $\hat{\bm{z}}$, respectively. The material derivative is denoted by $D_t = \partial_t + \bm{u} \cdot \nabla$. Now, we shall make the following three postulates.
\begin{enumerate}
\item \textbf{Linear background stratification:} there exists a motionless reference state where buoyancy $b$ is a linear increasing function on the vertical axis, $z$, and the pressure gradient is balanced with the buoyancy force. We then consider motion of disturbances superimposed on this reference state. The solutions are accordingly represented as
\begin{align*}
\bm{u} = \bm{u}', \quad b = N^2 z + b', \quad p = \frac{N^2 z^2}{2} + p' .
\end{align*}
\item \textbf{Linearization of equations:} disturbance components denoted with primes are sufficiently small, so that the material derivatives are approximated as $D_t \bm{u} \sim \partial_t \bm{u}'$ and $D_t b \sim \partial_t b' + N^2 w'$. \item \textbf{Traditional approximation:} the angular velocity $\bm{\Omega}$ in the Coriolis acceleration term is approximated by its vertical component, which is conventionally represented as $(f / 2) \hat{\bm{z}}$.
\end{enumerate}
Finally, rescaling the buoyancy perturbation as \begin{equation}b' = N \theta'\end{equation}
and dropping the primes on unknown variables, we obtain the model equations \eqref{eq:basic} with the divergence-free constraint $\nabla \cdot \bm{u} = 0$.}

\yo{\section{Eigenbundles of the bulk states} \label{app:eigenbundle}
\subsection{Inertia-gravity wave}
The solutions of the Fourier-transformed eigenvalue problem $\omega \hat{\bm{\psi}} = \bm{\mathcal{H}}_k \hat{\bm{\psi}}$ are obtained by hand. For the genuine solutions satisfying $\phi = 0$, the eigenvectors are represented using the Fourier coefficient of pressure, $\hat{p}$, in a single expression,
\begin{align} \label{eq:eigenvector}
\hat{\bm{\psi}} = \begin{pmatrix}
\hat{u} \\
\hat{v} \\
\hat{w} \\
\hat{\theta}
\end{pmatrix} 
= \hat{p} \begin{pmatrix}
\dfrac{k_x \omega + \ii k_y f}{\omega^2 - f^2} \\[12pt]
\dfrac{k_y \omega - \ii k_x f}{\omega^2 - f^2} \\[12pt]
\dfrac{- k_z \omega}{N^2 - \omega^2} \\[12pt]
\dfrac{\ii k_z N}{N^2 - \omega^2}
\end{pmatrix} .
\end{align}
It is convenient for the later consideration to write the complex pressure variable as $\hat{p} = \gamma \vert \hat{p} \vert$ and assume that $\vert \hat{p} \vert$ is determined from the normalization condition $\vert \hat{\bm{\psi}} \vert$ = 1. The argument factor, $\gamma \in U(1)$, represents the gauge freedom. To make explicit the parameter dependence, let us redefine the basic eigenvector, \eqref{eq:eigenvector}, as $\hat{\bm{\psi}} = \bm{\Psi}(k_x, k_y, \omega, \gamma)$. The eigenvectors for the inertia-gravity waves are then represented by $\bm{\Psi}(k_x, k_y, \omega_\pm, \gamma) \equiv \hat{\bm{\psi}}^\gamma_\pm$. Note that the solutions \eqref{eq:eigenvector} are invalid when $\omega = \pm f, \pm N$, and the eigenvectors of these singular cases are considered separately below.

In the following, we assume $f \neq 0$ so that the bulk spectrum has gaps at $- \vert f \vert < \omega < 0$ and $0 < \omega < \vert f \vert$. We define the Riemann sphere $S^2$ composed of $k \equiv k_x + \ii k_y \in \mathbb{C} \cup \{ \infty \}$, and inspect the eigenvector of the positive-frequency branch around points where $\omega^+ = \vert f \vert$ $(k = 0)$ or $\omega^+ = N$ $(k = \infty)$. In the limit of $k \to 0$, we learn $\hat{u} / \hat{p}, \hat{v} / \hat{p} \sim \mathcal{O}(k^{-1})$ while $\hat{w} / \hat{p}, \hat{\theta} / \hat{p} \sim \mathcal{O}(1)$, and accordingly
\begin{align*}
\lim_{k \to 0} \hat{\bm{\psi}}_+^\gamma = \frac{\gamma e^{\sign(f) \ii \alpha}}{\sqrt{2}} \begin{pmatrix}
1 \\
- \sign(f) \ii \\
0 \\
0
\end{pmatrix} ,
\end{align*}
where $\alpha$ denotes the argument of $k$, and $\sign(f) \equiv f / \vert f \vert$ is defined. This vector is single-valued at $k = 0$ by setting $\gamma=\gamma_\infty$ with
\begin{equation}
\gamma_\infty\equiv e^{- \sign(f) \ii \alpha} , \label{eq:gamma_infty} 
\end{equation}
in the same manner as the shallow-water system \cite{tauber2019bulk}. Next, in the limit of $k \to \infty$, we derive $\hat{u} / \hat{p}, \hat{v} / \hat{p} \sim \mathcal{O}(k)$ and $\hat{w} / \hat{p}, \hat{\theta} / \hat{p} \sim \mathcal{O}(k^2)$, and accordingly
\begin{align*}
\lim_{k \to \infty} \hat{\bm{\psi}}_+^\gamma = \frac{\sign(k_z)\gamma}{\sqrt{2}} \begin{pmatrix}
0 \\
0 \\
- 1 \\
\ii
\end{pmatrix} .
\end{align*}
Notably, in this limit the eigenvector apparently does not depend on the length or the angle of the horizontal wave vector, differing from the standard shallow water case \cite{delplace2017topological}. Physically, this motion is called the buoyancy oscillation in which the horizontal velocity vanishes. In this limit, the choice of $\gamma = \gamma_\infty$ defined in \eqref{eq:gamma_infty} does not produce a unique convergence of the eigenvector. Instead, fixing the gauge as $\gamma = \gamma_0$ with
\begin{align*}
\gamma_0 = 1 ,
\end{align*}
we may extend the range of the wave number to $k = \infty$.

Consequently, eigenvectors are definable on a compactified surface, $S^2$, and may compose a topologically non-trivial fiber bundle. For the visualization purpose, we shall specify the Riemann sphere by the standard stereographic projection: one-to-one mapping between $k \in \mathbb{C} \cup \{ \infty \}$ to $\bm{X} = (X_1, X_2, X_3) \in \mathbb{R}^3$ with $\vert \bm{X} - (0,0,1/2) \vert = 1/2$ via
\begin{align} \label{eq:stereographic}
k = \frac{X_1 + \ii X_2}{1 - X_3} ,
\end{align}
which is a homeomorphism of $S^2$. Figure~\ref{fig:phase_Riemann} demonstrates the bundle structure on $S^2$, which exhibits singularity in the polarization relation for any choice of gauge $\gamma$. This singularity is characterized by a topological invariant, namely, the Chern number.

\begin{figure}
  \centering
  \noindent\includegraphics[bb=0 0 1408 1025, width=\columnwidth]{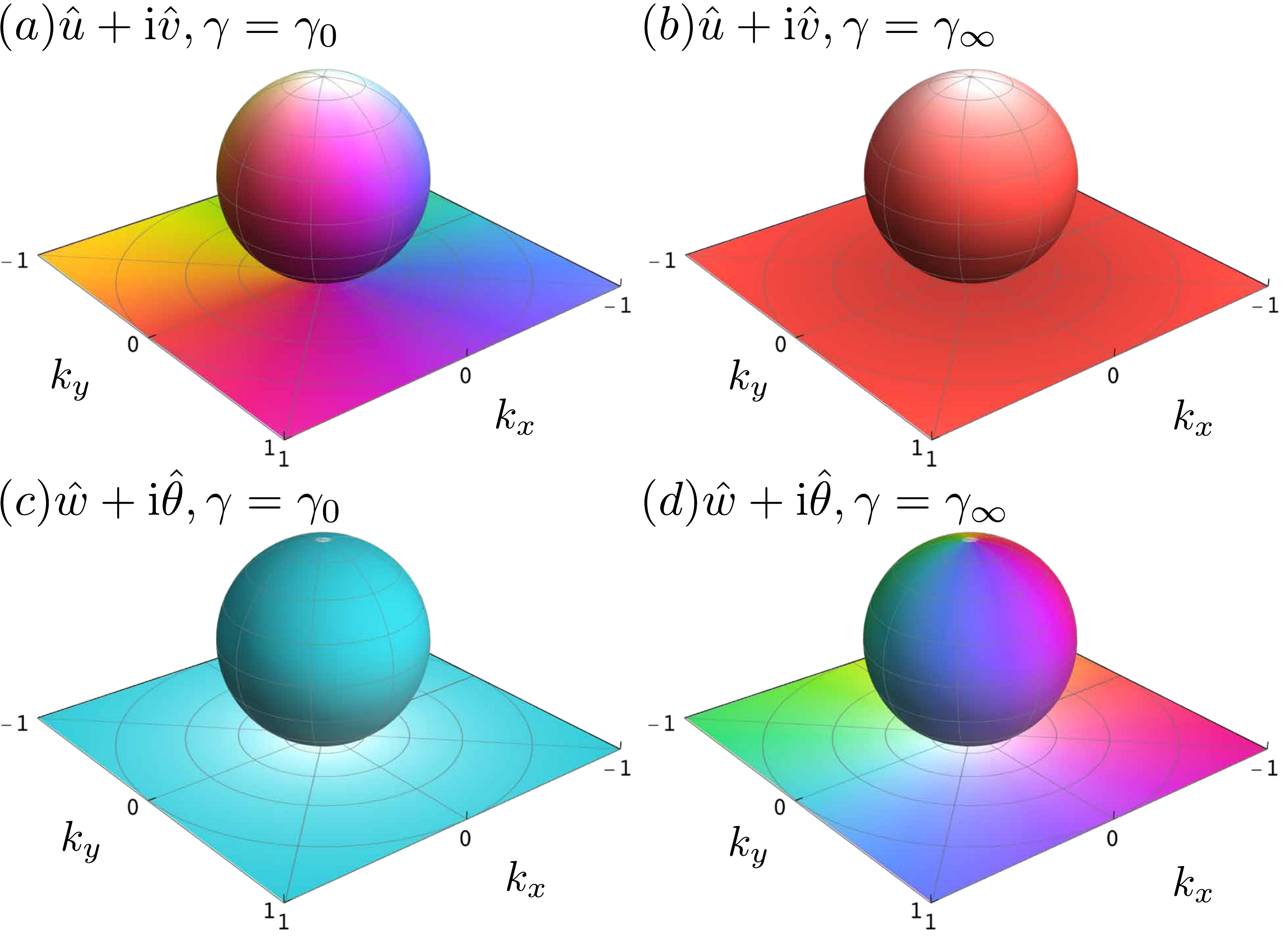}\\
  \caption{Bundle structure of the positive inertia-gravity wave band on the Riemann sphere composed of the horizontal wave number, $k = k_x + \ii k_y$. Parameters are $(k_z, f, N) = (2, 2, 5)$. Absolute values (shade) and phases (color) are shown for (a, b) $\hat{u} + \ii \hat{v}$, (c, d) $\hat{w} + \ii \hat{\theta}$. Depending on the choice of gauge $\gamma$, the location of the phase singularity changes. For $\hat{u}$ and $\hat{v}$, a singular point is located at $k = 0$ when $\gamma = \gamma_0$. For $\hat{w}$ and $\hat{\theta}$, a singular point is located at $k = \infty$ when $\gamma = \gamma_\infty$} \label{fig:phase_Riemann}
\end{figure}

To compute the Chern number, we first introduce the Berry connection \footnote{Sign definitions in the Coriolis parameter, forward/backward Fourier transforms, Berry connection, and the argument of a scattering coefficient may differ from those in other papers on the same topic. Specifically, the present definitions follow \cite{venaille2021wave} and \cite{venaille2023ray} but have differences from \cite{delplace2017topological}, \cite{tauber2020anomalous} and \cite{graf2021topology}.}, $\bm{A}_+^\gamma = \ii \langle \hat{\bm{\psi}}_+^\gamma, \nabla_k \hat{\bm{\psi}}_+^\gamma \rangle$, where $\nabla_k = (\partial_{k_x}, \partial_{k_y})$, and the brackets denote the inner products between complex vectors. The Berry connection defines the Berry curvature through $B_+ = \nabla_k \times \bm{A}_+^\gamma$, where $\bm{a} \times \bm{b} = a_1 b_2 - a_2 b_1$ for $\bm{a} = (a_1, a_2)$ and $\bm{b} = (b_1, b_2)$ is understood. The Berry curvature is further related to the Chern number via $\mathcal{C} = (1 / 2\pi) \int_{S^2} B_+ dk_x dk_y$, which is the $U(1)$-bundle form of \eqref{eq:chern}. To process further, let us write the two representations of polarization relations as $\hat{\bm{\psi}}^\infty_+$ for $\gamma = \gamma_\infty$ and $\hat{\bm{\psi}}^0_+$ for $\gamma = \gamma_0$, which are defined in $S^2 \setminus \{ \infty \}$ and $S^2 \setminus \{ 0 \}$, respectively. We also define the corresponding Berry connections as $\bm{A}_+^\infty$ and $\bm{A}_+^0$. From Stokes' theorem, the Chern number is computed as
\begin{align*}
\mathcal{C}_+ & = \frac{1}{2 \pi} \int_{S^2} B_+ dk_x dk_y = \frac{1}{2 \pi} \oint \left( \bm{A}_+^\infty - \bm{A}_+^0 \right) \cdot d\bm{s} \\
& = \frac{\ii}{2 \pi} \oint \left( \gamma_\infty^\dag \nabla_k \gamma_\infty - \gamma_0^\dag \nabla_k \gamma_0 \right) \cdot d\bm{s} \\
& = \frac{\ii}{2 \pi} \oint d \log \frac{\gamma_\infty}{\gamma_0} = \sign(f) ,
\end{align*}
in which the integration contour of the line integrals surrounds $k = 0$ in the counterclockwise direction. We have used the fact that $\gamma$ is unitary when deriving the second line. For the negative frequency branch, $\omega^-$, the Chern number flips its sign so that we may write $\mathcal{C}_{\pm} = \pm \sign(f)$.

\subsection{Zero-frequency mode}
For degenerate zero eigenvalues, the corresponding eigenspace is spanned by the following two eigenvectors:
\begin{subequations} \label{eq:zero_eigenvector}
\begin{align}
\frac{1}{\sqrt{N^2 (k_x^2 + k_y^2) + f^2 k_z^2}} \begin{pmatrix}
N k_y \\
- N k_x \\
0 \\
- f k_z
\end{pmatrix} &
\equiv \hat{\bm{\psi}}_{0 g}  \\
\frac{1}{\sqrt{k_x^2 + k_y ^2 + k_z^2}} \begin{pmatrix}
k_x \\
k_y \\
k_z \\
0
\end{pmatrix} &
\equiv \hat{\bm{\psi}}_{0 a} ,
\end{align}
\end{subequations}
which correspond to the geostrophic flow and the auxiliary field, respectively. Because any linear combination of \eqref{eq:zero_eigenvector} is also an eigenvector, the set of normalized eigenvectors can be generally represented by
\begin{align*}
\begin{pmatrix}
\hat{\bm{\psi}}^\Gamma_{0+} \\[5pt]
\hat{\bm{\psi}}^\Gamma_{0-}
\end{pmatrix} = \Gamma \begin{pmatrix}
\hat{\bm{\psi}}_{0 g} \\[5pt]
\hat{\bm{\psi}}_{0 a}
\end{pmatrix} ,
\end{align*}
where $\Gamma$ is an arbitrary $2 \times 2$ matrix that belongs to $U(2)$. Note that the two vectors in \eqref{eq:zero_eigenvector} cannot be separately defined at $k = \infty$. However, if we set
\begin{align*}
\Gamma = \frac{1}{\sqrt{2} \vert k \vert} \begin{pmatrix}
k & \ii k \\
k^\dag & - \ii k^\dag
\end{pmatrix} ,
\end{align*}
the transformed eigenvectors converge as
\begin{align*}
\lim_{k \to \infty} \hat{\bm{\psi}}^\Gamma_{0 \pm} = \frac{1}{\sqrt{2}} \begin{pmatrix}
\pm \ii \\
-1 \\
0 \\
0
\end{pmatrix} .
\end{align*}
Therefore, this eigenspace composes a fiber bundle with a structure group of $U(2)$ over a closed surface $S^2$. Also in this case, the Berry connection and the Berry curvature derive from $\hat{\bm{\psi}}^\Gamma_{0\pm}$, but now comprise $2 \times 2$ elements. We refer to Chapter 3.6.4 of \cite{vanderbilt2018berry} for detailed expressions. Anyway, due to the universal formula, $\sum_n \mathcal{C}_n = 0$, we learn that the Chern number of the middle band, $\mathcal{C}_0$, vanishes.
}

\yo{\section{Spectrum of $\mathcal{H}$ around an equator} \label{app:equator}
The present estimates of the Chern numbers \eqref{eq:chern_number} differ from previous reports for a rotating shallow-water model. Because the standard shallow water equation does not admit unique eigenvectors in the large wave number limit, it is necessary to add the odd-viscosity terms parameterized by a coefficient $\epsilon$ for a regularization purpose \cite{tauber2019bulk,souslov2019topological}. The resulting Chern numbers were either $\{ \pm 2, 0, \mp 2 \}$ or $\{ 0, 0, 0 \}$ dependent on the signs of $f$ and $\epsilon$. Although the two models apparently exhibit inconsistency, they have a common property. In any case, Chern numbers vary when $f$ changes signs, and the differences are represented by
\begin{align} \label{eq:difference_chern_number}
\{ \Delta \mathcal{C}_-, \Delta \mathcal{C}_0, \Delta \mathcal{C}_+ \} = \{ - 2, 0, 2 \} .
\end{align}
This result predicts the two kinds of unidirectional modes trapped around the equator that fill the frequency gaps between the geostrophic flow and the inertia-gravity wave bands, which surely correspond to the equatorial Kelvin and Yanai waves. The dispersion relations for these two waves in an elementary stratified fluid model on the equatorial $\beta$-plane under the hydrostatic approximation were analytically derived by Matsuno \cite{matsuno1966quasi}. In the following, we extend it to a non-hydrostatic problem. As a side note, similar expressions including compressibility are derived in~\cite{qian2003nonhydrostatic}.

We shall set $f = \beta y$ in \eqref{eq:basic} and impose the boundary condition such that the variables vanish at $y \to \pm \infty$. Making an ansatz, $(\bm{u}, \theta, p) = (\hat{\bm{u}}, \hat{\theta}, \hat{p}) e^{\ii (k_x x + k_z z - \omega t)}$, we obtain a set of ordinary differential equations,
\begin{subequations} \label{eq:beta-governing}
\begin{align}
- \ii \omega \hat{u} & = \beta y \hat{v} - \ii k_x \hat{p} \\
- \ii \omega \hat{v} & = - \beta y \hat{u} - \hat{p}_y \label{eq:beta-v} \\
- \ii \omega \hat{w} & = N \hat{\theta} - \ii k_z \hat{p} \label{eq:beta-w} \\
- \ii \omega \hat{\theta} & = - N \hat{w} \label{eq:beta-theta} \\
0 & = \ii k_x \hat{u} + \hat{v}_y + \ii k_z \hat{w} , \label{eq:beta-cont}
\end{align}    
\end{subequations}
which determine the eigenvalue $\omega$ and the corresponding meridional structure functions. Here, we have used a subscript to represent a differentiation. It is convenient to solve \eqref{eq:beta-w} and \eqref{eq:beta-theta} to write $\hat{w}$ as a function of $\hat{p}$, and insert it into \eqref{eq:beta-cont}, to derive an equation system holding three variables,
\begin{align*}
- \ii \omega \hat{u} - \beta y \hat{v} + \ii k_x \hat{p} & = 0 \\
- \ii \omega \hat{v} + \beta y \hat{u} + \hat{p}_y & = 0 \\
\ii k_x \hat{u}+ \hat{v}_y - \frac{\ii k_z^2 \omega \hat{p}}{N^2 - \omega^2} & = 0 .
\end{align*}
These expressions are almost the same as those resulting from the hydrostatic model. Therefore, the analytical solutions are obtained in a similar fashion. The crucial step is to write down a single equation with respect to $\hat{v}$ as
\begin{align*}
\hat{v}_{yy} + \left( \frac{k_z^2 \omega^2}{N^2 - \omega^2} - k_x^2 - \frac{k_x \beta}{\omega} - \frac{k_z^2 \beta^2 y^2}{N^2 - \omega^2} \right) \hat{v} = 0 .
\end{align*}
Notably, when $\omega^2 > N^2$, solutions are oscillatory in the limit of $y \to \pm \infty$, so that $\omega$ does not have discrete eigenvalues in this range. Because we are interested in equator-trapped eigenmodes, $\omega^2 < N^2$ is assumed in the following. Let us introduce a characteristic length scale as $L = \left[(N^2 - \omega^2) / (k_z^2 \beta^2) \right]^{1/4}$ and define the dimensionless coordinate by $Y = y / L$. The equation becomes $\hat{v}_{YY} + (F - Y^2) \hat{v} = 0$ with
\begin{align*}
F = \left( \frac{N^2 - \omega^2}{k_z^2 \beta^2} \right)^{1/2} \left( \frac{k_z^2 \omega^2}{N^2 - \omega^2} - k_x^2 - \frac{k_x \beta}{\omega} \right) .
\end{align*}
From the analogy with a quantum harmonic oscillator, the eigenvalues are determined by $F = 2 n + 1$ with $n = 0, 1, 2, \ldots$.
We shall write this equation in polynomial form,}
\begin{widetext}
\yo{\begin{align}
& (k_x^2 + k_z^2)^2 \omega^6 + 2 k_x (k_x^2 + k_z^2) \beta \omega^5 + (- 2 k_x^2 (k_x^2 + k_z^2) N^2 + (k_x^2 + k_z^2 (2n + 1)^2) \beta^2 ) \omega^4 \nonumber \\
& - 2 k_x (2k_x^2 + k_z^2) N^2 \beta \omega^3 + (k_x^4 N^4 - (2k_x^2 + k_z^2 (2n + 1)^2) N^2 \beta^2) \omega^2 + 2 k_x^3 N^4 \beta \omega + k_x^2 N^4 \beta^2 = 0 , \label{eq:beta-dispersion}
\end{align}}
\end{widetext}
\yo{for which roots satisfying $F > 0$ are meaningful.

In the special case of $n = 0$, \eqref{eq:beta-dispersion} is factorized to}
\begin{widetext}
\yo{\begin{align} \label{eq:Yanai}
\left( (k_x^2 + k_z^2) \omega^2 - k_x^2 N^2 \right) \left( (k_x^2 + k_z^2) \omega^4 + 2 k_x \beta \omega^3 + (- k_x^2  N^2 + \beta^2) \omega^2 - 2 k_x N^2 \beta \omega - N^2 \beta^2 \right) = 0 .
\end{align}}
\end{widetext}
\yo{The second factor yields the Yanai-wave solution. The first factor as well as $F = 1$ yields
\begin{align*}
\omega = - \frac{N k_x}{\sqrt{k_x^2 + k_z^2}} , 
\end{align*}
which is not a proper solution. If we adopt this dispersion relation, the structure function of $\hat{p}$ determined by
\begin{align*}
( - \beta^2 y^2 + \omega^2 ) \hat{v} + \ii (k_x \beta y + \omega \partial_y) \hat{p} = 0 .
\end{align*}
cannot satisfy the boundary conditions, $\lim_{y \to \pm \infty} \hat{p} = 0$.

The Kelvin wave solution is obtained by setting $\hat{v} = 0$ to assume the geostrophic balance between $\hat{u}$ and $\hat{p}$ in \eqref{eq:beta-v}. The remaining equations in \eqref{eq:beta-governing} are algebraically solved to yield
\begin{align} \label{eq:equator-Kelvin}
\omega = \frac{N k_x}{\sqrt{k_x^2 + k_z^2}} .
\end{align}
We note that another solution, $\omega = - N k_x / \sqrt{k_x^2 + k_z^2}$ is prohibited because it yields exponential divergence of $\hat{u}$ and $\hat{p}$ in the limit of $y \to \pm \infty$.

In summary, the whole set of dispersion relations of equator-trapped modes are specified by \eqref{eq:beta-dispersion}, \eqref{eq:Yanai}, and \eqref{eq:equator-Kelvin}. The algebraic equations are solved numerically to create Fig.~\ref{fig:spectrum_beta}. In the same way as reported for shallow-water cases \cite{delplace2017topological,tauber2019bulk}, the dispersion curves of the Kelvin and Yanai waves connect the separated frequency bands in agreement with \eqref{eq:difference_chern_number}.

\begin{figure}[t]
  \centering
  \noindent\includegraphics[bb=0 0 372 380, width=0.6\columnwidth]{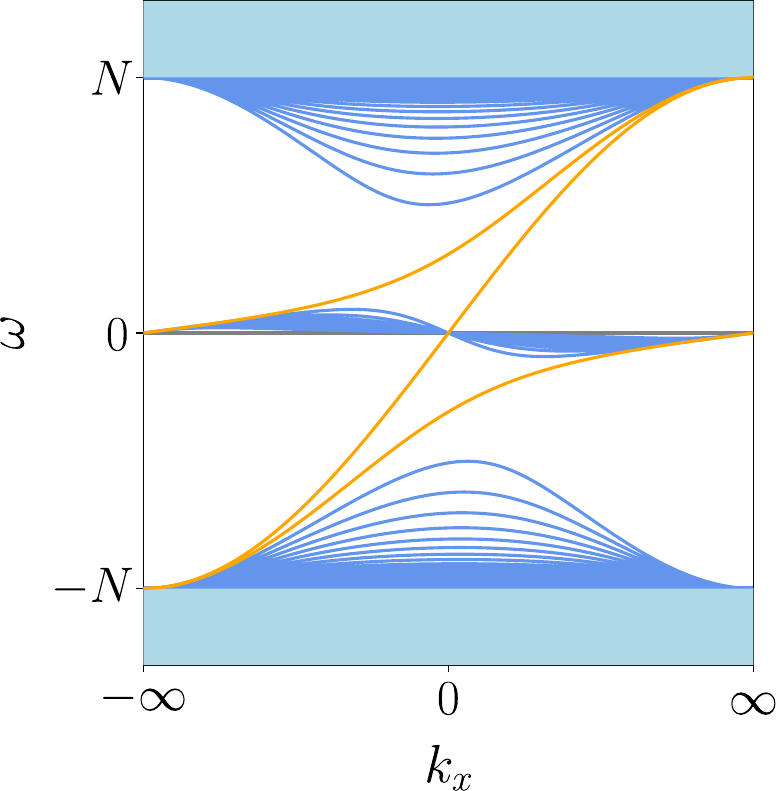}\\
  \caption{\yo{The spectrum of $\mathcal{H}$ for an equatorial beta plane $f = \beta y$. Parameters are $(k_z, \beta, N) = (2, 1, 5)$. The orange curves represent the Kelvin and Yanai waves.}} \label{fig:spectrum_beta}
\end{figure}
}

\yo{
\section{Edge-mode dispersion relations} \label{app:edge_mode}
\subsection{Derivation of the edge mode on an elastic wall}
Edge-mode solutions trapped around the wall at $y = 0$ are generally represented as
\begin{align}
\bm{\psi} = \hat{\bm{\Psi}} (k_x, \kappa, \omega, \gamma) e^{\ii (k_x x + \kappa y + k_z z - \omega t)} ,
\end{align}
where $\kappa$ is an imaginary number satisfying $\Im \kappa > 0$ and related to $\omega$ via \eqref{eq:kappa}.
The reduced form of the elastic boundary condition \eqref{eq:boundary_y0} turns into $\hat{v} = \ii a \omega \hat{p}$. Making use of \eqref{eq:eigenvector}, we obtain
\begin{align} \label{eq:edge_k_omega}
\frac{\kappa \omega - \ii k_x f}{\omega^2 - f^2} = \ii a \omega .
\end{align}
Eliminating $\kappa$ from \eqref{eq:kappa} and \eqref{eq:edge_k_omega}, we derive an algebraic equation determining $\omega$ as
\begin{widetext}
\begin{align} \label{eq:edge-mode-dispersion}
a^2 \omega^6 - (a^2 f^2 + a^2 N^2) \omega^4 + 2 a k_x f \omega^3 - (k_x^2 + k_z^2 - a^2 N^2 f^2) \omega^2- 2 a N^2 k_x f \omega + N^2 k_x^2 = 0 .
\end{align}
\end{widetext}
Roots of this expression satisfying $\Im \kappa > 0$ with \eqref{eq:edge_k_omega} yield the dispersion relations of the edge modes. Several numerical results are demonstrated in Fig.~\ref{fig:edge_modes}.

\subsection{Asymptotic limits}
The algebraic equation \eqref{eq:edge-mode-dispersion} cannot generally be solved by hand. Here, we instead inspect the asymptotic behavior of the solutions in several special cases.

First, when $a = 0$ is set, \eqref{eq:edge-mode-dispersion} reduces to a simple quadratic form,
\begin{align}
(k_x^2 + k_z^2) \omega^2 - N^2 k_x^2 = 0 ,
\end{align}
coinciding with the dispersion relation of a non-rotating internal gravity wave. Condition $\Im \kappa > 0$ allows a brunch satisfying $\omega / k > 0$. These are well-known features of a coastal Kelvin wave trapped on a rigid wall.

Next, in the limit of $a \to \infty$, \eqref{eq:edge-mode-dispersion} turns into a fully factorized form,
\begin{align}
\omega^2 (\omega^2 - f^2) (\omega^2 - N^2) = 0.
\end{align}
Therefore, we learn that an edge-mode frequency approaches either $\omega=0, \pm f$ or $\pm N$, among which only the inertial-oscillation mode, $\omega = \pm f$, is isolated from the bulk spectrum. In this limit, the boundary condition becomes $p = 0$, i.e., fluid elements on the boundary can move without any pressure force from the wall. In a shallow-water system, the same situation arises if water depth abruptly becomes infinity at a boundary. Also in that case, the inertial oscillation mode is identified as a trapped solution \cite{iga1995transition}.

Finally, we consider the modification of the Kelvin-wave dispersion relation for a finite $a$. It is verified from \eqref{eq:edge-mode-dispersion} that $(k_x, \omega) = (0, 0)$ is always a solution. Therefore, there should exist a non-dispersive mode in the long-wavelength limit, namely, the long Kelvin wave influenced by the wall elasticity. To derive its phase speed, we insert $\omega = c k_x$ into \eqref{eq:edge-mode-dispersion} and take the limit of $k_x \to 0$ to have
\begin{align} \label{eq:boundary_kelvin_1}
(a^2 N^2 f^2 - k_z^2) c^2 - 2 a N^2 f c + N^2 = 0 .
\end{align} 
Because of the condition $\Im \kappa \geq 0$, when $f > 0$, a solution with $1 / (a f)> c > 0$ is allowed. Besides, since the left-hand side of \eqref{eq:boundary_kelvin_1} changes signs between $c=0$ and $c=1 / (a f)$, there always exists a unique solution of $c$ in this range. We thus learn that existence of the long Kelvin wave is robust, and its propagation direction remains the same for any~$a$.

\subsection{Origin of the high-frequency edge modes} \label{app:high_frequency}
As demonstrated in Fig.~\ref{fig:edge_modes}, an edge mode with the frequency greater than $N$ exists when $a > 0$. Here, we analytically elucidate how this mode emerges from wall elasticity. For this purpose, we shall recover the finite mass of springs and rewrite the boundary condition \eqref{eq:bc_general} as
\begin{align*}
\frac{\partial_t^2 v}{\omega_s^2} + v = - a \partial_t p \quad \text{at} \quad y = 0 .
\end{align*}
Accordingly, in the derivation of the edge-mode dispersion relations, we should replace \eqref{eq:edge_k_omega} with
\begin{align} \label{eq:edge_finite_mass}
\left( - \frac{\omega^2}{\omega_s^2} + 1 \right) \frac{\kappa \omega - \ii k_x f}{\omega^2 - f^2} = \ii a \omega .
\end{align}
To make the discussion simpler, we shall regard $\omega$ and $\omega_s$ as large and $a$ as small, i.e., considering the asymptotic limit of $0 < a \ll \vert f \vert, N, k_x, k_z \ll \omega_s \sim \omega$. Then, \eqref{eq:kappa} gives the estimate of the meridional wave number as
\begin{align*}
\kappa \sim \ii \sqrt{k_x^2 + k_z^2} + \mathcal{O} \left( \frac{1}{\omega_s^2} \right) .
\end{align*}
Bearing this in mind, we expand \eqref{eq:edge_finite_mass} asymptotically to obtain
\begin{align} \label{eq:freq_edge_finite_mass}
\frac{1}{\omega^2} = \frac{1}{\omega_s^2} + \frac{\ii a}{\kappa} + \mathcal{O} \left( \frac{1}{\omega_s^3} \right) .
\end{align}
The positive root is
\begin{align*}
\omega \sim \sqrt{\frac{1}{1 / \omega_s^2 + a / \sqrt{k_x^2 + k_z^2}}} .
\end{align*}
In the limit of $a \to 0$, as expected, we obtain the free oscillation mode of the elastic boundary, $\omega = \omega_s$. On the other hand, in the limit of zero mass $\omega_s \to \infty$, we obtain $\omega \sim \sqrt[4]{k_x^2 + k_z^2} / \sqrt{a}$. This solution connects $(k_x, \omega) = (\pm \infty, \infty)$ and is hence consistent with the numerical solutions in Fig.~\ref{fig:edge_modes}.

To give the physical interpretation to the present result, it is important to notice that the mass of fluid influenced by this edge mode is proportional to the $e$-folding scale in the meridional direction, $\ii / \kappa$. The restoring coefficient that pushes and pulls this fluid mass is specified by $\lambda = 1 / a$. Therefore, from basic knowledge of harmonic oscillation, one may define the natural frequency of this oscillatory motion as $\omega_f \equiv \sqrt{\kappa / (\ii a)}$. We consequently understand that the squared inverse of the edge-mode frequency, $1 / \omega^2$ in \eqref{eq:freq_edge_finite_mass}, results as the sum of those of two oscillators, $1 / \omega^2_s$ and $1 / \omega^2_f$. Their restoring forces commonly originate from the elasticity of the wall and does not dependent on $f$ or $N$. Note that this simplification is valid in the asymptotic limit; if the frequency is not large, effects from the buoyancy and Coriolis force cannot be neglected.
}

\yo{
\section{Scattering coefficient} \label{app:scatter}
This Section presents a detailed construction of the scattering coefficient $S$ as well as a series of paths $C_\delta$ on which $S$ is computed. First, it should be kept in mind that the expression of $S$ depends on the choice of gauge. We then have to choose $\gamma$ such that $C_\delta$ encloses the singular point of $\hat{\bm{\psi}}^\gamma_+(k_x, k_y)$ in the upper-half plane of $\mathbb{C}$ composed of $k = k_x + \ii k_y$. When $f > 0$, we can arbitrarily move the location of a singular point to $k = \zeta$ by choosing a gauge as
\begin{align*}
\gamma = \sqrt{\frac{1 - \zeta / k}{1 - (\zeta / k)^\dag}} .
\end{align*}
This study sets $\zeta = \ii$, as denoted by crosses in Fig.~\ref{fig:edge_modes}f, throughout the analysis. Applying the boundary condition \eqref{eq:boundary_y0} to $\bm{\psi} = \hat{\bm{\psi}}^\gamma_+(k_x, -\kappa) e^{\ii (k_x x - \kappa y + k_z z - \omega_+ t)} + S(k_x, \kappa) \hat{\bm{\psi}}^\gamma_+(k_x, \kappa) e^{\ii (k_x x + \kappa y + k_z z - \omega_+ t)}$, we derive the explicit form of $S$ as
\begin{gather*}
S(k_x, \kappa) = - \frac{g(k_x, - \kappa)}{g(k_x, \kappa)} \quad \text{with} \quad \\
g(k_x, \kappa) \equiv \left[ \frac{\kappa \omega_+ - \ii k_x f}{\omega_+^2 - f^2} - \ii a \omega_+ \right] \sqrt{\frac{1 - \zeta / (k_x + \ii \kappa)}{1 - \zeta^\dag / (k_x - \ii \kappa)}} ,
\end{gather*}
where $\omega_+$ is also a function $k_x$ and $\kappa$.

To define a path $C_\delta$ in the $k$-plane, let us recall the stereographic projection from the Riemann sphere, \eqref{eq:stereographic}. We assign a set of coordinates, $(\delta, \mu)$, to represent the sphere in the Cartesian coordinates as
\begin{align*}
X_1 & = \sqrt{1/4 - \delta^2} \sin \mu \\
X_2 & = \delta \\
X_3 & = 1/2 - \sqrt{1/4 - \delta^2} \cos \mu .
\end{align*}
It is now clear that $C_\delta$ corresponds to a circle that is the intersection of the sphere and a plane of $X_2 = \delta$. Location on the path is specified by $\mu$, which denotes the angle around the center of the circle measured from the bottom (Fig.~\ref{fig:edge_modes}f). The corresponding wave numbers are derived as
\begin{align*}
k_x & = \frac{X_1}{1 - X_3} = \frac{\sqrt{1/4 - \delta^2} \sin \mu}{1 / 2 + \sqrt{1 / 4 - \delta^2} \cos \mu} \\
k_y & = \frac{X_2}{1 - X_3} = \frac{\delta}{1 / 2 + \sqrt{1 / 4 - \delta^2} \cos \mu} ,
\end{align*}
which are in the same form as described in the Fig.~\ref{fig:edge_modes}  caption. We insert these expressions into $S(k_x, k_y)$ to create Fig.~\ref{fig:edge_modes}g-i. In the computation of the logarithmic function, the Riemann surface is chosen such that $\ii \log S$ is continuous for $\mu \in (0, 2\pi)$.
}

\yo{
\section{Bulk-edge correspondence under a different boundary condition} \label{app:different_bc}
This section tests the formula of the bulk-edge correspondence, i.e., \eqref{eq:Levinson}, under a boundary condition other than the elastic wall. For this purpose, we consider a condition of 
\begin{align} \label{eq:differ-bc}
v = a \partial_x p \quad \text{at} \quad y = 0 .
\end{align}
For any real $a$, \eqref{eq:differ-bc} holds the criteria (i) of the Hermitian property discussed in Section~\ref{sec:wall}. Note that this boundary condition is here introduced as a mathematical toy model.

\subsection{Dispersion relations of edge modes}
In the present case, in the derivation process of the edge-mode dispersion relations elucidated in Appendix~\ref{app:edge_mode}, condition \eqref{eq:edge_k_omega} should be replaced by
\begin{align} \label{eq:differ-k-omega}
\frac{\kappa \omega - \ii k_x f}{\omega^2 - f^2} = \ii a k_x .
\end{align}
Combining this with \eqref{eq:kappa}, after some algebra, we derive a quadratic equation in terms of $\omega^2$,
\begin{align} \label{eq:differ-edge-equation}
a^2 k_x^2 \omega^4 - & \left[ (a f - 1)^2 k_x^2 + a^2 k_x^2 N^2 + k_z^2 \right] \omega^2 \nonumber \\
+ & (af - 1)^2 k_x^2 N^2 = 0 .
\end{align}
The left-hand side is not negative when $\omega^2 = 0$, and its determinant is
\begin{widetext}
\begin{align*}
D = & \left[ (a f - 1)^2 k_x^2 + a^2 k_x^2 N^2 + k_z^2 \right]^2 - 4 a^2 (af - 1)^2 k_x^4 N^2 \\
= & \left[ (a f - 1)^2 k_x^2 - a^2 k_x^2 N^2\right]^2 + 2 \left[ (a f - 1)^2 k_x^2 + a^2 k_x^2 N^2\right] k_z^2 + k_z^4 > 0 .
\end{align*}
\end{widetext}
Consequently, $\omega^2$ always has a pair of positive roots. It is informative to write down the possible four roots of $\omega$ as
\begin{align*}
\omega = \pm \sqrt{\frac{\left[ (a f - 1)^2 k_x^2 + a^2 k_x^2 N^2 + k_z^2 \right] \pm \sqrt{D}}{2 a^2 k_x^2}} .
\end{align*}
In light of the condition $\Im \kappa > 0$ and \eqref{eq:differ-k-omega}, we always obtain two meaningful solutions of $\omega$ for any $a \neq 0$ and $k_x$.

\begin{figure*}
  \noindent\includegraphics[width=0.9\textwidth, bb=0 0 543 213]{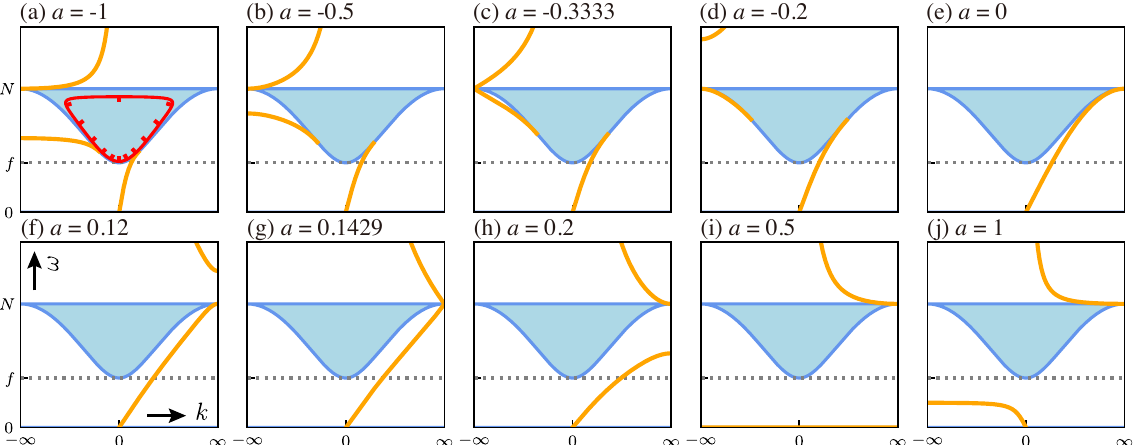}\\
  \caption{Dispersion curves of edge modes (orange) for cases of a new boundary condition, $v = a \partial_x p$ at $y = 0$, are superimposed on the bulk spectrum (cyan). In (a), the red curve denotes a path $C_\delta$ with $\delta = 0.2$ in the same way as Fig.~3b. Parameters are $(k_z, f, N) = (2, 2, 5)$.} \label{fig:edge_different_bc}
\end{figure*}

\begin{figure*}
  \noindent\includegraphics[width=0.9\textwidth, bb=0 0 558 193]{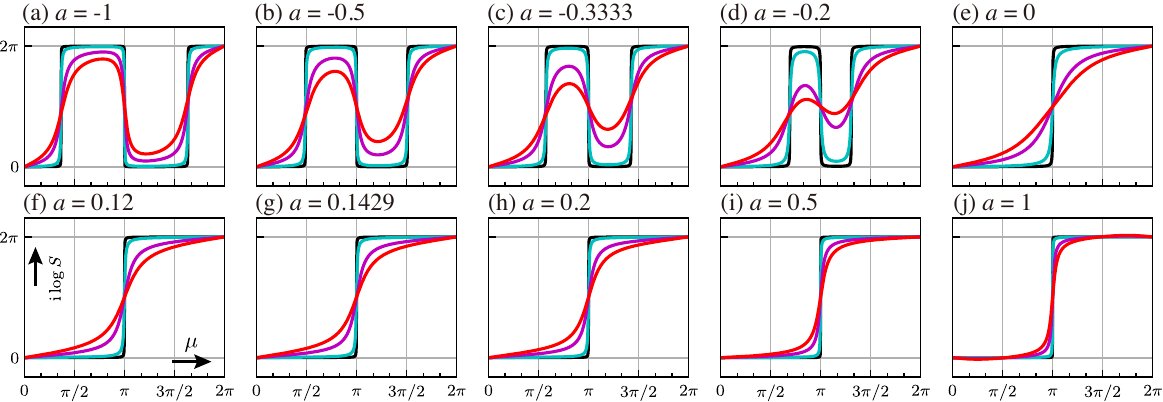}\\
  \caption{Plots of phases in the scattering coefficient $S$ for positive inertia-gravity waves with the boundary condition of $v = a \partial_x p$ at $y = 0$, computed along $C_\delta$ for various $a$ as corresponding to Fig.~\ref{fig:edge_different_bc}. Parameter $\delta \in (0.001, 0.01, 0.1, 0.2)$ is distinguished by colors in the same manner as Fig.~\ref{fig:edge_modes}.} \label{fig:scattering_different_bc}
\end{figure*}

\subsection{Major properties in edge-mode dispersion curves}
Also in this problem, it is informative to inspect the characters of dispersion curves for some special choices of $k_x$ and $a$. First, when $k_x \to 0$, $\omega$ approaches either $0$ or $\pm \infty$. In these two limits, \eqref{eq:differ-edge-equation} is approximated by
\begin{align*}
& - k_z^2 \omega^2 + (a f - 1)^2 k_x^2 N^2 = 0 \\
\text{and} \quad & a^2 k_x^2 \omega^4 - k_z^2 \omega^2 = 0 ,
\end{align*}
respectively. We therefore have four roots,
\begin{align*}
\omega \sim \pm \frac{(af - 1) N k_x}{k_z}, \pm \frac{k_z}{a k_x} ,
\end{align*}
for which the condition of $\Im \kappa > 0$ permits only one sign in each set of double signs. Notably, there always exists a uni-directional non-dispersive wave in the long-wavelength limit.

Next, when $k_x \to \pm \infty$, \eqref{eq:differ-edge-equation} is approximated by
\begin{align*}
a^2 \omega^4 - \left[ (a f - 1)^2 + a^2 N^2 \right] \omega^2 + (af - 1)^2 N^2 = 0 ,
\end{align*}
which is factorized to
\begin{align*}
(a^2 \omega^2 - (af - 1)^2) (\omega^2 - N^2) = 0 .
\end{align*}
Remarkably, for any choice of $a$, either $\omega = N$ or $\omega = -N$ is always a solution. The other pair of roots are generally apart from the bulk spectrum and discontinuous between $k_x = \infty$ and $k_x = - \infty$, which is not surprising as the boundary condition \eqref{eq:differ-bc} is apparently singular at $k_x = \pm \infty$. More interestingly, these two roots dependent on $a$ coincide with the other fixed roots at $\omega = \pm N$ if
\begin{align*}
a = \frac{1}{N + f} , - \frac{1}{N - f} .
\end{align*}
When $a$ passes either of these two values, two edge modes with $\vert \omega \vert > N$ and $\vert \omega \vert < N$ are reconnected.

When $a = 0$, the boundary condition coincides with the simplest rigid condition, and there exists a unique Kelvin-wave solution. In the limit of $a \to \pm \infty$, \eqref{eq:differ-edge-equation} approaches
\begin{align*}
\omega^4 - (f^2 + N^2) \omega^2 + f^2 N^2 = 0 ,
\end{align*}
whose solutions are $\omega = \pm f, \pm N$. This result is equivalent to that from the elastic boundary condition.

Finally, when $a = 1/f$, two roots with the absolute values smaller than $N$ reduce to $\omega = 0$. As $a$ passes this value, the originally Kelvin-like mode changes its propagation direction, and will approach the inertial oscillation. All of these properties of dispersion curves are visualized in Fig.~\ref{fig:edge_different_bc}.

\subsection{Scattering coefficient}
The scattering coefficient $S$ is computed along a series of paths $C_\delta$ in the same manner as Appendix~\ref{app:scatter} by redefining a function $g$ such that
\begin{gather*}
g(k_x, \kappa) \equiv \left[ \frac{\kappa \omega_+ - \ii k_x f}{\omega_+^2 - f^2} - \ii a k_x \right] \sqrt{\frac{1 - \zeta / (k_x + \ii \kappa)}{1 - \zeta^\dag / (k_x - \ii \kappa)}} .
\end{gather*}
The results are demonstrated in Fig.~\ref{fig:scattering_different_bc}. As expected, in every case, a phase jump in $S$ corresponds to a connection point of an edge-mode dispersion curve to the bulk spectrum. Notably, reconnection at $k = \pm \infty$ occurs keeping the net count of connection points. We thus verify that the bulk-edge correspondence in a form of Levinson's theorem \eqref{eq:Levinson} is established even under a different boundary condition.
}


\bibliography{rotating_stratified_fluid}

\end{document}